\documentclass[preprint, 12pt]{elsarticle}

\usepackage{graphicx}
\usepackage{tabularx}
\usepackage{mathtools}
\usepackage[version=4]{mhchem}
\usepackage[dvipsnames]{xcolor}
\usepackage{multirow}
\usepackage[colorlinks=true]{hyperref}

\newcolumntype{L}[1]{>{\raggedright\let\newline\\\arraybackslash\hspace{0pt}}m{#1}}
\newcolumntype{C}[1]{>{\centering\let\newline\\\arraybackslash\hspace{0pt}}m{#1}}
\newcolumntype{R}[1]{>{\raggedleft\let\newline\\\arraybackslash\hspace{0pt}}m{#1}}

\def\mntwe{\texorpdfstring{M\MakeLowercase{n}\textsubscript{12}}{Mn12}}
\def\mostwo{\texorpdfstring{M\MakeLowercase{o}S\textsubscript{2}}{MoS2}}

\def\compone{{\texorpdfstring{M\MakeLowercase{n}\textsubscript{12}-H}{Mn12-H}}}
\def\comptwo{{\texorpdfstring{M\MakeLowercase{n}\textsubscript{12}-CH\textsubscript{3}}{Mn12-CH3}}}
\def\compthr{{\texorpdfstring{M\MakeLowercase{n}\textsubscript{12}-CHCl\textsubscript{2}}{Mn12-CHCl2}}}
\def\compfou{{\texorpdfstring{M\MakeLowercase{n}\textsubscript{12}-C\textsubscript{6}H\textsubscript{5}}{Mn12-C6H5}}}

\def\ligone{{\texorpdfstring{-H}{-H}}}
\def\ligtwo{{\texorpdfstring{-CH\textsubscript{3}}{-CH3}}}
\def\ligthr{{\texorpdfstring{-CHCl\textsubscript{2}}{-CHCl2}}}
\def\ligfou{{\texorpdfstring{-C\textsubscript{6}H\textsubscript{5}}{-C6H5}}}

\def\eV{\,\textrm{eV}}
\def\V{\,\textrm{V}}

\begin{document}


\title{Tuning the Magnetic Anisotropy Energy of \mostwo{}-supported \mntwe{} complexes by Electric Field: A First-Principles Study} 

\author[aff1]{Shuanglong Liu}

\author[aff2]{Adam V. Bruce}

\author[aff2]{Dmitry Skachkov\fnref{fn2}}

\author[aff1]{James N. Fry}

\author[aff1]{Hai-Ping Cheng\corref{cor1}}

\address[aff1]{Department of Physics, Northeastern University, Boston, Massachusetts 02115, USA} 

\address[aff2]{Department of Physics, Quantum Theory Project, and Center for Molecular Magnetic Quantum Materials, University of Florida, Gainesville, Florida 32611, USA} 

\cortext[cor1]{ha.cheng@northeastern.edu}

\fntext[fn2]{Current address: University of Central Florida, Orlando, FL 32816 USA}


\begin{abstract}
In this work, we examine low-energy adsorption configurations of four dodecanuclear manganese single-molecule magnets \ce{[Mn12O12(O2CR)16(H2O)4]} (\mntwe{}), where the ligand R being H, \ce{CH3}, \ce{CHCl2} or \ce{C6H5}, on a molybdenum disulfide (\mostwo{}) monolayer using force field and density functional theory calculations. The van der Waals interaction is shown to be crucial for determining the adsorption energy. Some electrons transfer from the substrate to the molecules upon surface adsorption, resulting in a reduction of the magnetic anisotropy energy of \mntwe{}. Since the lowest unoccupied molecular orbital of \mntwe{} is close to the valence band of \mostwo{}, a negative electric field is more effective in modulating charge transfer and energy band alignment, and thus altering the magnetic anisotropy energy, compared with a positive electric field. A significant increase in the magnetic anisotropy energy of \mntwe{} with the ligand \ce{R=CHCl2} or \ce{R=C6H5} under a sufficiently high electric field has been predicted. Our calculations show that the molecules remain intact on the surface both before and after the electric field is applied. Finally, a two-level system formed by different adsorption configurations is evaluated, and the tunability of its energy barrier under an electric field is demonstrated. Our study sheds light on tuning the properties of single-molecule magnets using an electric field, when the molecules are supported on a surface. 
\end{abstract}

\maketitle

\section{Introduction}
\label{sect:introduction}

Relativity manifests itself as energy splitting of spin multiplets in transition metal complexes under zero magnetic field.
The splitting of the lowest-energy spin manifold can be probed by electron paramagnetic resonance spectroscopy~\cite{Sessoli1993-Mn12-JACS, Hill2004-EPR-Mn12, Barra2008-EPR-Mn12} and inelastic electron tunneling spectroscopy~\cite{Xue2008-IETS-CoPc,Vincent2012-IETS-TbPc2}. 
Given a total spin $S \ge 1$, the magnetic molecule will have an axial spin or magnetic easy axis if the ground state predominantly consists of $m_S = \pm S$ after zero-field splitting (ZFS). 
Such molecules are referred to as single-molecule magnets (SMMs). The first identified SMM is \ce{[Mn12O12(O2CR)16(H2O)4]}~\cite{Sessoli1993-Mn12-JACS} with \ce{R=CH3}, \comptwo{}, which is a member of the family of polynuclear transition metal SMMs~\cite{Barra1999-Fe4,Murrie2010-CoII-SMM}. Different families of SMMs were identified later on, including 3d single-ion magnets~\cite{Murrie2015-SIM} and lanthanide-based SMMs~\cite{Woodruff2013-Ln-SMM}.
SMMs feature quantized magnetic hysteresis~\cite{Sessoli1993-Mn12-Nature, Novak1995-Mn12}, quantum tunneling of magnetization~\cite{Friedman1996-QTM,Gatteschi2003-QTM}, and slow magnetic relaxation~\cite{Sessoli1993-Mn12-Nature}.
The slow magnetic relaxation enables the application of SMMs as miniature units for classical information storage.
Applications of SMMs in quantum information processing have also been proposed and pursued~\cite{Leuenberger2001-QIStorage,Meier2003-QComputing,Hill2025-Qubit-SMM}.

Employing SMMs in planar electronic devices inevitably involves deposition of SMMs on a surface. 
\comptwo{} and its derivatives, which all share the magnetic core \ce{Mn12O12}, have been deposited on metal~\cite{Kahle2011-IonBeam-hBN,Mannini2008-XMCD,Cornia2003-FuncMn12,Voss2007-FuncGold,Coronado2005-FuncGold,Moroni2008-PVS,Saywell2010-Mn12onGold}, semi-metal~\cite{Sun2013-STM-Bi111,Art2020-Mn12onGr}, semiconductor~\cite{Condorelli2004-FuncSi001}, and insulator~\cite{Clemente-Leon1998-LBFilm,Cavallini2005-DVDMold,Laskowska2019-Mn12-NPs,Pastukh2021-SilicaNP,Martínez2007-FuncSiO2,Moroni2008-PVS,Ruiz‐Molina2003-Mn12onPolymer} substrates. 
SMM properties have been observed in thin films~\cite{Clemente-Leon1998-LBFilm,Ruiz‐Molina2003-Mn12onPolymer,Moroni2008-PVS} on flat surfaces and in monolayers~\cite{Pastukh2021-SilicaNP} on the surface of nanoparticles. In the latter case, an ensemble of nanoparticles was used in magnetometry measurements; otherwise, the signal from a single monolayer of \mntwe{} would be too weak to detect.
In thin films, the top few layers of \mntwe{} may lose their SMM properties, which can be probed by X-ray magnetic circular dichroism (XMCD) measurements.~\cite{Sessoli2010-XMCDReview} 
A submonolayer of \mntwe{} molecules on a gold surface was found to lose its SMM properties, according to XMCD measurements.~\cite{Mannini2008-XMCD} 
This is because a large portion of \ce{Mn^{3+}} and \ce{Mn^{4+}} ions in the magnetic core are reduced to \ce{Mn^{2+}} ions.~\cite{Voss2007-FuncGold} 
Surface functionalization~\cite{Voss2007-FuncGold} and the insertion of an \textit{h}-BN monolayer~\cite{Kahle2011-IonBeam-hBN} were found to be effective in mitigating the reduction of \mntwe{} molecules. 
It is noteworthy that \mntwe{} molecules are not sublimable due to their thermal instability. Suitable deposition techniques for \mntwe{} molecules were discussed in the review by Gabarr\'{o}-Riera \textit{et al.}~\cite{Gabarro-Riera2023-SMMonSurface}

The effective energy barrier for tunneling of magnetization can be estimated by the magnetic anisotropy energy (MAE) in \textit{ab initio} calculations~\cite{Mark1999-Mn12-MAE}, where MAE is defined as the energy difference between the state where the spin is along the easy axis and the state where the spin is perpendicular to the easy axis. Previously, the MAE of \mntwe{} with different ligands adsorbed on graphene was calculated~\cite{Xiangguo2014-Mn12-MAE,Shuanglong2022-Mn12-MAE}, and the MAE of \compone{} on graphene was found to be reduced by electrostatic doping. However, the reduction of MAE is unfavorable for data storage applications due to faster magnetization relaxation. 
In this work, we investigate the surface adsorption of four \mntwe{} complexes, i.e., \compone{}, \comptwo{}, \compthr{}, and \compfou{}, on a \mostwo{} monolayer. 
The semiconducting nature of \mostwo{} limits the electron transfer to/from \mntwe{}, which helps to preserve the SMM properties; and the layered form of \mostwo{} permits application in low-dimensional electronic and spintropic devices. 
%
%
It is found that the MAE of \compthr{} and \compfou{} on \mostwo{} can be significantly enhanced by an electric field. In contrast, the MAE of the other two \mntwe{} complexes becomes smaller after surface adsorption, both with and without the electric field. Charge and band analyses were performed to understand the change in MAE. 
In addition, possible transitions between different adsorption configurations are demonstrated by a two-level system in \compone{}/\mostwo{}. Such transitions could lead to dynamical magnetization for sparsely distributed \mntwe{} molecules on the surface, which may be used to generate random numbers for stochastic computing~\cite{Wang2024-StochasticComputing}.

%

\section{Method}
\label{sect:method}

Density functional theory~\cite{Kohn1964-HKTheorems,Kohn1965-KSEquations} (DFT) was used to relax the atomistic structures and calculate the magnetic anisotropy energy. Structural relaxations were performed using the Vienna Ab initio Simulation Package~\cite{Kresse1995-AIMD,Kresse1999-US2PAW} (VASP), and the magnetic anisotropy energy was calculated using the SIESTA package~\cite{Solar2002-SIESTA}. 
Prior to DFT structural relaxation, the universal force field as implemented in the Gaussion package~\cite{Frisch2016-GaussianPackage} was first applied to find candidate adsorption configurations. This helps save computational costs, given the large system size and the high number of possible adsorption configurations. 
The search for a two-level system in \mntwe{}/\mostwo{} was based on the method of climbing image nudged elastic band (cNEB) as implemented in VASP.~\cite{Henkelman2000-cNEB} 

In VASP calculations, the energy cutoff for plane waves was set to $500 \eV$. The Perdew-Burke-Ernzerhof exchange correlation energy functional~\cite{Perdew1996-PBE-XC} together with the projector augmented-wave pseudopotentials~\cite{Bochl1994-PAW} were used. Only the $\Gamma$ point was sampled in reciprocal space given the large supercell size ($a = b > 29$ \AA{}, $c \ge 35\,$\AA{}). The van der Waals interaction was included via the DFT+D3 method~\cite{Grimme2010-DFT-D3}. The energy and force tolerances were set to $1\times10^{-6} \eV$ and $0.005\eV/$\AA{} respectively. The force tolerance was increased to $0.02 \eV/$\AA{} in cNEB calculations. 
In the SIESTA calculations, double-$\zeta$ polarized basis functions were used for all atoms except C and H atoms, for which single-$\zeta$ polarized basis functions were used. A mesh cutoff of $200\,\textrm{Ry}$ was applied for the real-space grid. Norm-conserving pseudopotentials, as generated by the Troullier-Martins scheme~\cite{Troullier1991-TM-Pseudopotential}, were employed. The same exchange-correlation energy functional was used as in the VASP calculations. The reciprocal space was sampled by a $16 \times 16 \times 1$ Monkhorst-Pack grid~\cite{Pack1977-MonkhorstPack-k-mesh} to calculate the MAE accurately. More $k$-points are affordable in SIESTA calculations due to the use of localized basis sets. The smearing parameter was set to $4.3 \, \textrm{meV}$. 
The spin-orbit coupling strength parameter was set to be 1.34 in MAE calculations~\cite{Shuanglong2022-Mn12-MAE}.

The following procedure was used in the force field calculations: First, the convex hull of a chosen \mntwe{} molecule was found using the Python package Scipy~\cite{2020SciPy-NMeth}. Second, for every face of the convex hull, the \mntwe{} molecule was placed on \mostwo{} with the face parallel to the \mostwo{} substrate. Third, the force-field-based total energy was minimized using the basin hopping algorithm under the following constraints: a) The substrate is fixed; b) The molecule is treated as a rigid body; and c) The chosen face is kept parallel to the substrate. Lastly, all local energy minima were collected and compared to find the lowest energy configuration. At least 15,000 adsorption configurations were obtained for each \mntwe{} complex after constrained structural relaxation. The lowest-energy adsorption configuration, as determined by the force field, was taken as the initial geometry for structural relaxation by DFT. One or more initial geometries were prepared manually for DFT structural relaxation, depending on the system size. The magnetic easy axis of each \mntwe{} complex was made perpendicular to the two-dimensional substrate in one of the manually prepared initial geometries. The number of contacts between the molecule and the substrate was maximized in other manually prepared initial geometries. Some adsorption configurations of \compone{} on \mostwo{} were identified as local energy minima along the transition pathway during cNEB calculations.

\section{Results}
\label{sect:results}

\begin{figure}[htb!]
\centering
\includegraphics[width=1.0\columnwidth]{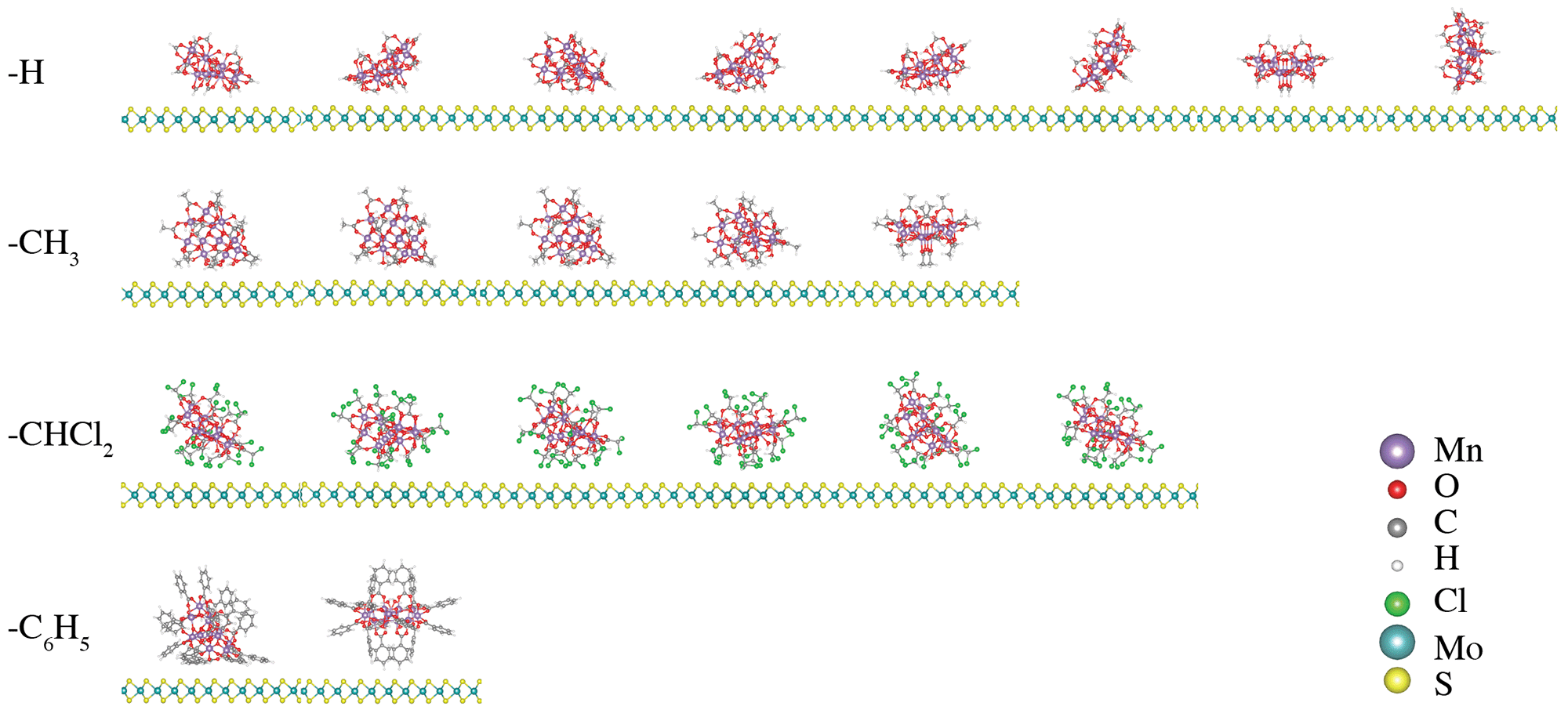}
\caption{\label{fig:structure}Relaxed adsorption configurations of \mntwe{} molecules with different ligands on \mostwo{} according to DFT calculations. The adsorption energy increases for the configurations from left to right.}
\end{figure}

\begin{figure}[htb!]
\centering
\includegraphics[width=0.6\columnwidth]{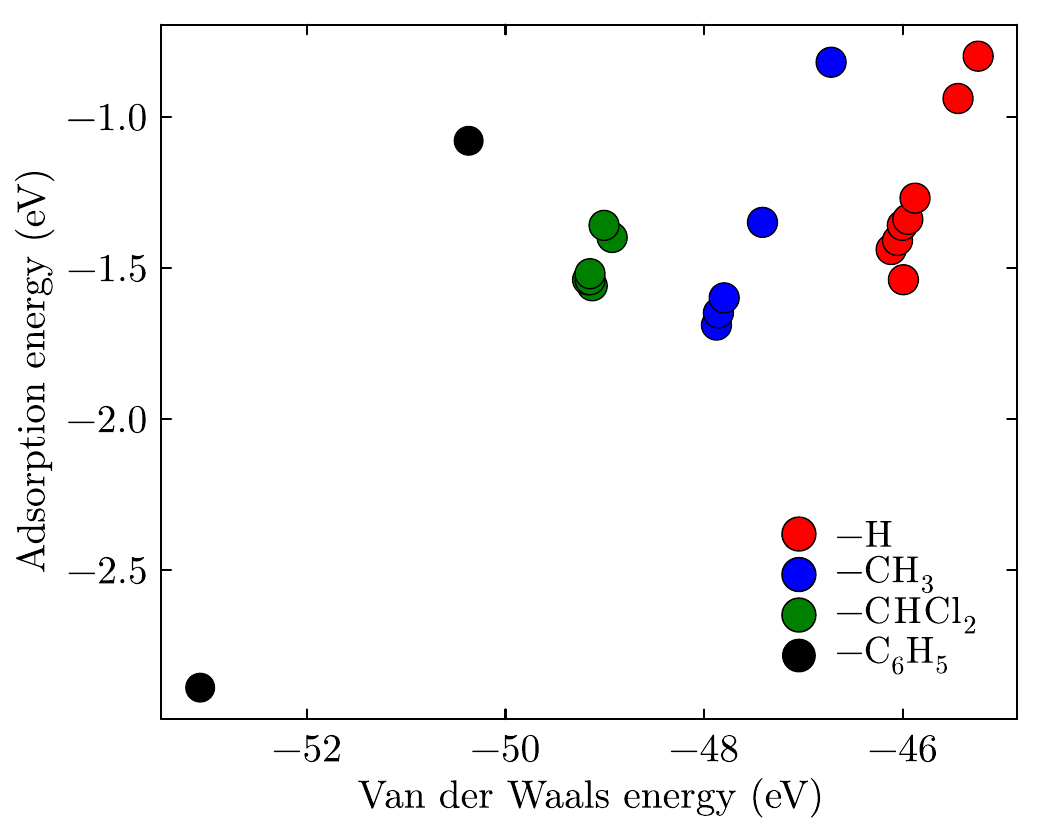}
\caption{\label{fig:vdw}Adsorption energy versus van der Waals energy for various adsorption configurations of \mntwe{} molecules with different ligands on \mostwo{}.}
\end{figure}

Fig.~\ref{fig:structure} shows the adsorption configurations for the \mntwe{} complexes on \mostwo{} as relaxed by DFT. 
The adsorption configurations are arranged in order of adsorption energy, with the leftmost configuration being the most stable for each \mntwe{} complex. 
Fig.~\ref{fig:vdw} shows the adsorption energy versus van der Waals energy for all configurations.
There is a clear trend in which the adsorption energy increases with the van der Waals energy. 
Across different adsorption configurations, the adsorption energy varies by 0.74, 0.87, 0.20, and $1.81 \eV$ for \compone{}, \comptwo{}, \compthr{}, and \compfou{}, respectively. 
Qualitatively, adsorption configurations with more contacts with the substrate are more stable. This is reasonable since the molecule-substrate interaction is dominated by van der Waals forces. 
In the most stable adsorption configuration, the distance between the \mntwe{} molecule and the \mostwo{} substrate is 1.98, 2.23, 2.94, and 2.53 \AA{} for  \compone{}, \comptwo{}, \compthr{}, and \compfou{}, respectively, as measured from the average plane of the top-layer S atoms to the nearest atom of the \mntwe{} molecule. The correlation between the molecule-substrate distance and the adsorption energy is weak.
For the most stable adsorption configurations, the adsorption energies are ordered as follows: $E_\mathrm{ads}$(\compfou{}) $<$ $E_\mathrm{ads}$(\comptwo{}) $<$ $E_\mathrm{ads}$(\compthr{}) $\sim$ $E_\mathrm{ads}$(\compone{}). 
The adsorption energy for \compfou{} is significantly lower (by $1 \eV$ or more) than that of the other three \mntwe{} complexes because the four \ligfou{} ligands in contact with the substrate become almost parallel to it, enhancing the molecule-substrate interaction.

\begin{figure}[htb!]
\centering
\includegraphics[width=0.6\columnwidth]{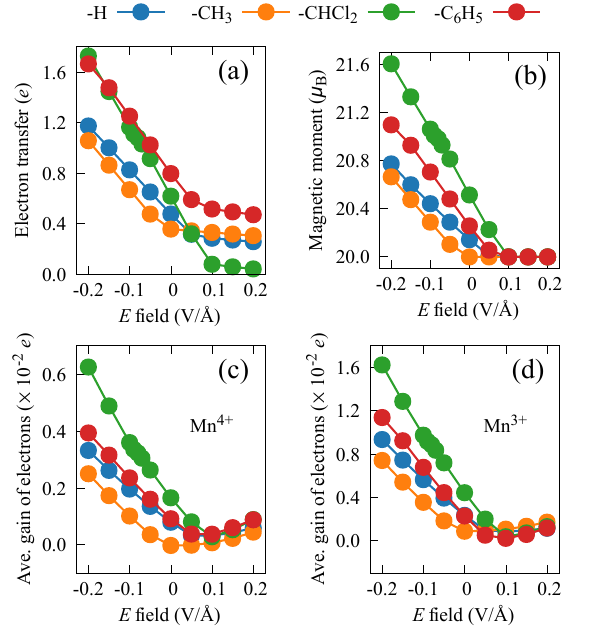}
\caption{\label{fig:transfer}(a) Electron transfer from \mostwo{} to \mntwe{} and (b) total magnetic moment versus electric field. (c) Average gain of electrons per \ce{Mn^{4+}} ion and (d) \ce{Mn^{3+}} ion versus electric field.} 
\end{figure}

\begin{figure}[htb!]
\centering
\includegraphics[width=1.0\columnwidth]{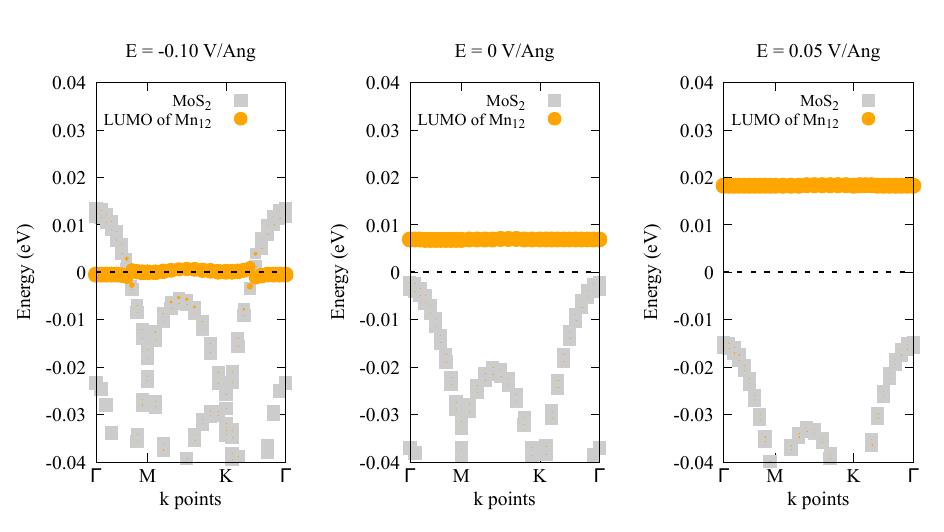}
\caption{\label{fig:bands}Energy bands of the hetero-structure \mntwe{}-H/\mostwo{}. The size of squares (circles) is proportional to the projected density states of \mostwo{} (\mntwe{}).}
\end{figure}

Mulliken charge analysis was carried out using the lowest energy adsorption configuration for each \mntwe{} complex, as shown in Fig.~\ref{fig:transfer}a. 
At zero electric field, \compone{}, \comptwo{}, \compthr{}, and \compfou{} gain 0.476, 0.356, 0.617, and 0.796 electrons from \mostwo{}, respectively. 
The charge transfer indicates that the lowest unoccupied molecular orbital of \mntwe{} is close to the valence band of \mostwo{}, which is confirmed by a plot of the energy bands for \compone{}/\mostwo{} in Fig.~\ref{fig:bands}. 
Most of the transferred electrons reside on the ligands of \mntwe{}, only 0.022, 0.007, 0.042, and 0.022 electrons migrate onto the \ce{Mn^{3+}} and \ce{Mn^{4+}} ions for  \compone{}, \comptwo{}, \compthr{}, and \compfou{}, respectively. 
Interestingly, each \ce{Mn^{4+}} of the central \ce{Mn4O4} cube loses 0.0003 electrons on average for \comptwo{}/\mostwo{} under zero electric field. 
For all other \mntwe{} complexes on \mostwo{} under zero electric field, both \ce{Mn^{3+}} and \ce{Mn^{4+}} ions gain electrons, as shown in Figs.~\ref{fig:transfer}c and \ref{fig:transfer}d. 
Fig.~\ref{fig:transfer}b shows that the magnetic moment of the heterostructures exceeds $20 \,\mu_B$, which is the magnetic moment of an isolated \mntwe{} molecule. 
This implies that the lowest unoccupied molecular orbital is dominated by the spin majority channel for all the \mntwe{} complexes considered here. 
The \mostwo{} substrate remains non-magnetic upon the adsorption of magnetic molecules. 
When a positive electric field, directed from the substrate to the molecule, is applied, electrons are pulled back to the substrate, leading to a reduction of the magnetic moment. Accordingly, the lowest unoccupied molecular orbital of \mntwe{} is lifted and shifts away from the valence band of \mostwo{}, as shown in Fig.~\ref{fig:bands}c. 
The magnetic moment of the heterostructures reaches $20 \,\mu_B$ at $ 0.1 \V/$\AA{} and remains at this value for electric fields between 0.1 and $0.2 
\V/$\AA{}. This is because the highest occupied molecular orbital of \mntwe{} is still far below the conduction band of \mostwo{}. 
When a negative electric field is applied, more electrons are transferred from the \mostwo{} substrate to the \mntwe{} molecules. 
It is noteworthy that \compthr{} and \compfou{} gain significantly more electrons than \compone{} and \comptwo{}. 
This is reasonable, as the electron-accepting ability of the \ligthr{} and \ligfou{} ligands is stronger than that of the \ligone{} and \ligtwo{} ligands.
A negative electric field can induce an inversion between the lowest unoccupied molecular orbital of \mntwe{} and the valence band of \mostwo{}, as observed in the band structure of Fig.~\ref{fig:bands}a for \compone{}/\mostwo{} under an electric field of $-0.1\ V/$\AA{}.
At an electric field of $-0.2\ V/$\AA{}, the electron transfer to \compthr{} and \compfou{} exceeds 1.6 electrons. 
%
%
It was known that an additional electron prefers to localize around a Mn atom when \mntwe{} is reduced.~\cite{Skachkov2022-ReducedMn12} 
However, the transferred electrons are delocalized over the whole molecule rather than localized in this case. This is because the molecular structure is fixed at the zero-field geometry. 
Structural relaxation for \compthr{}/\mostwo{} under $-0.2$ V/\AA{} did not result in the localization of the transferred electrons either, implying that the Mulliken charge analysis may have overestimated the electron transfer.

\begin{table}[htb!]
    \centering
    \begin{tabular}{>{\raggedleft\arraybackslash}p{2.5cm} >{\centering\arraybackslash}p{2cm} >{\centering\arraybackslash}p{2cm} >{\centering\arraybackslash}p{2cm} >{\centering\arraybackslash}p{2cm}}
    \hline
    & -H & -$\textrm{CH}_{3}$ & -$\textrm{CHCl}_{2}$ & {-$\textrm{C}_{6}\textrm{H}_{5}$}     \\
    \hline
    Free \mntwe{}                       & $5.999$     & $6.128$    & $6.415$      & {$6.190$}  \\ 
    \multirow{2}{*}{Isolated \mntwe{}}  & $5.621$     & $5.905$    & $5.726$      & $6.153$    \\ \noalign{\vskip -3pt} 
                                        & ($-6.3\%$)  & ($-3.6\%$) & $(-10.7\%)$  & ($-0.6\%$) \\ 
    \multirow{2}{*}{\mntwe{}/\mostwo{}} & $5.329$     & $6.052$    & $5.014$      & $5.672$    \\ \noalign{\vskip -3pt} 
                                        & ($-11.2\%$) & ($-1.2\%$) & $(-21.8\%)$  & ($-8.4\%$) \\ 
    \hline
    \end{tabular}
    \caption{\label{table:maeads}Magnetic anisotropy energy (in meV) of free isolated \mntwe{} molecules, isolated \mntwe{} molecules as relaxed on \mntwe{}, and \mntwe{} molecules adsorbed on \mostwo{}. The values in the parentheses represent the relative change compared to the free molecules.} 
\end{table}

The \mntwe{} molecules feature a magnetic easy axis that is perpendicular to the average plane of the eight \ce{Mn^{3+}} ions and also perpendicular to the top and bottom faces of the central \ce{Mn4O8} cube. 
Since the \ce{Mn4O8} cube is less distorted compared with the \ce{Mn^{3+}} ions when a \mntwe{} molecule is adsorbed on \mostwo{}, the central \ce{Mn4O8} cube was used to determine the magnetic easy axis. 
%
%
Table~\ref{table:maeads} shows the MAE for the \mntwe{} complexes before and after surface adsorption. Compared to the free \mntwe{} complexes, the MAE decreases by 1.2--21.8\% after surface adsorption, depending on the ligand. 
The change in MAE is due to both the structural change in the molecule and the electronic structure change, the latter of which involves electron transfer from the substrate and band alignment with the substrate. To elucidate the effects of the electronic structure change, the MAE was calculated for the \mntwe{} molecules relaxed on the substrate, in the absence of the substrate. The results are also shown in Table~\ref{table:maeads}. 
Electronic structure changes account for about half of the MAE variation in \compone{} and \compthr{}. For \compfou{}, this same mechanism is responsible for most of the MAE change. As for \comptwo{}, the electronic structure change tends to increase the MAE, while the structural change reduces it, with the structural change having a stronger effect in this case.

\begin{figure}[htb!]
\centering
\includegraphics[width=0.5\columnwidth]{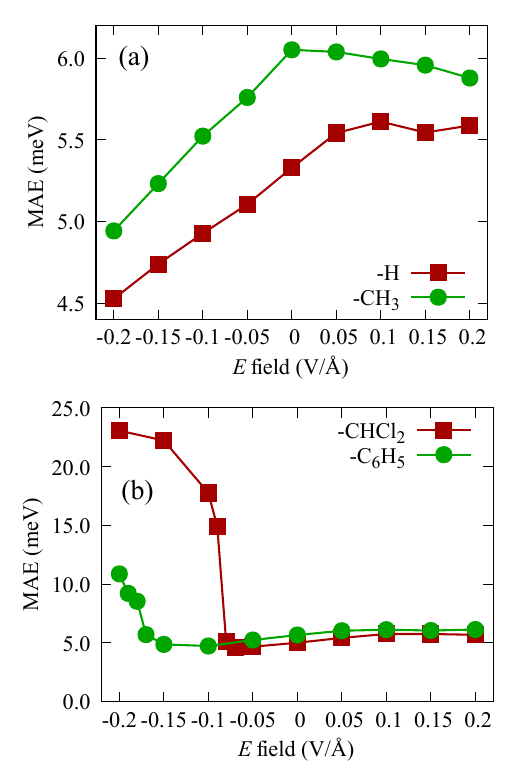}
\caption{\label{fig:mae}Magnetic anisotropy energy of \mntwe{} molecules with different ligands adsorbed on a \mostwo{} monolayer versus the electric field.}
\end{figure}

Fig.~\ref{fig:mae} shows the calculated MAE of \mntwe{} adsorbed on \mostwo{} as a function of the applied vertical electric field. 
Overall, the MAE is more sensitive to a negative electric field than to a positive electric field. 
When an electric field of $+0.05\V/$\AA{} is applied, the MAE of \compone{}, \compthr{}, and \compfou{} increases, while that of \comptwo{} decreases compared to its value at zero electric field. 
The increase in MAE for the former three complexes is caused by the sizable charge transfer from the \mntwe{} molecules back to the \mostwo{} substrate as shown in Fig.~\ref{fig:transfer}a. 
The decrease in MAE for the latter complex is likely caused by the change in the band alignment since the charge transfer is negligible. 
%
%
When a negative electric field is applied, the MAE of \compone{} and \comptwo{} consistently decreases as the field strength increases, whereas the MAE of \compthr{} and \compfou{} initially decreases before rising. The MAE starts to increase with the strength of electric field at $-0.07\V/$\AA{} for \compthr{} and $-0.1 \V/$\AA{} for \compfou{}. 
At $-0.2\V/$\AA{}, the MAE of \compone{}, \comptwo{}, \compthr{}, and \compfou{} was tuned by $-15.1\%$, $-18.3\%$, $+360.0\%$, and $+91.7\%$, respectively, compared to its value at zero electric field. 
The bigger change in MAE under negative electric fields is due to greater electron transfer and energy band inversion that were presented earlier. 
The increase in the MAE of \compthr{} and \compfou{} on \mostwo{} under a negative electric field is sharp. This cannot be solely attributed to the increase in electron transfer, which is smooth. Changes in the band alignment are likely responsible for the sharp increase in the MAE.
Further study is needed to gain a quantitative understanding of the sharp increase in MAE for \compthr{} and \compfou{} on \mostwo{}.

The above MAE results are for a frozen molecular structure that was relaxed under zero electric field.
To know the effect of structural relaxation under finite electric fields, the structure of \compthr{}/\mostwo{} was further relaxed in the presence of external electric field. 
The molecular structure remains intact after relaxation under electric fields. 
Table~\ref{table:maerel} shows the MAE of fully relaxed \compthr{} on \mostwo{} under selected electric fields, as well as that of the frozen structure.
First, the overall trend in the change of MAE remains the same. Second, the over-threefold enhancement of MAE at $-0.2\V/$\AA{} remains valid for the fully relaxed structure. Third, a significant change in MAE due to structural change may occur at certain electric fields, such as $-0.1 \V/$\AA{}.

\begin{table}[htb!]
    \centering
    \begin{tabular}{ccc}
    \hline
    $E$ field (V/{\AA}) & \hphantom{1pt} Frozen \hphantom{1pt} & \hphantom{1pt} Optimized \hphantom{1pt} \\
    \hline
    $-0.20$   & $23.066$  & $23.670$ \\
    $-0.10$   & $17.762$  & $10.403$ \\
    $~0.10$   & $~5.763$  & $~5.715$ \\
    $~0.20$   & $~5.689$  & $~4.788$ \\
    \hline
    \end{tabular}
    \caption{\label{table:maerel}Magnetic anisotropy energy (in meV) of \mntwe{}-\ce{CHCl2} adsorbed on a \ce{MoS2} monolayer with and without structural relaxation under different electric fields. The frozen structure is identical to the structure relaxed under zero electric field.} 
\end{table}

\begin{figure}[htb!]
\centering
\includegraphics[width=0.9\columnwidth]{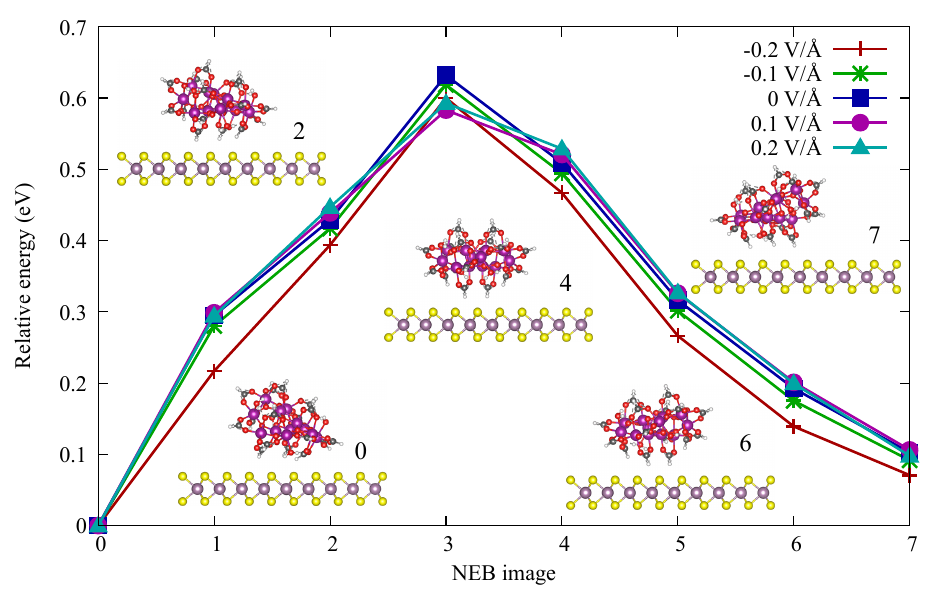}
\caption{\label{fig:twolevel}The minimum energy path for a \mntwe{}-H molecule adsorbed on a \ce{MoS2} monolayer under zero and finite electric fields.}
\end{figure}

Two-level systems may form between the stable and metastable adsorption configurations of \mntwe{} on \mostwo{}. A proof-of-concept calculation was performed for \compone{}/\mostwo{} to demonstrate such a two-level system and its tunability by an electric field.
Fig.~\ref{fig:twolevel} shows the minimal energy path for \compone{}/\mostwo{} under different electric fields. 
The energy of the initial configuration is set to zero for all electric fields to facilitate the comparison of energy barriers.
The energy barrier at zero electric field is $0.63 \eV$ and decreases under both positive and negative electric fields. The maximum reduction in the energy barrier is $0.05 \eV$. 
As seen from the insets of Fig.~\ref{fig:twolevel}, the molecule wobbles from one side to the other along the transition pathway. 
The energy barrier is relatively high compared to room temperature since the number of contacts is at least 3 for both the initial and final states, whereas the number of contacts is only two for the transition state.
The energy barrier may be lower for two-level systems with a transition state that has more contacts between the molecule and the substrate. 
With a lower energy barrier, the out-of-plane magnetic moment is more likely to change as the adsorption configuration changes, which is favorable for generating random numbers based on the magnetic moment. 

\section{Conclusion}
\label{sect:conclusion}

In conclusion, the van der Waals interaction is important for determining the stable adsorption configurations of \mntwe{} complexes on \mostwo{}. Ligand bending at the interface strongly enhances the adsorption energy for \compfou{}/\mostwo{}. The \mntwe{} molecules remain intact after surface adsorption, both in the presence and absence of an external electric field. The MAE of all four \mntwe{} complexes on \mostwo{} under zero electric field is noticeably smaller than that of isolated molecules. The MAE is more sensitive to a negative electric field, which promotes electron transfer and causes band inversion, than to a positive electric field. Our calculations show that the MAE can be enhanced using an electric field by a factor of 4.6 and 1.9, respectively, for \compthr{} and \compfou{} on \mostwo{}. Further study is needed to gain a deeper understanding of such dramatic increase in MAE. The MAE of \compone{} and \comptwo{} on \mostwo{} decreases significantly with a negative electric field. Lastly, a two-level system formed by different adsorption configurations was identified, demonstrating the possibility of a wobbling motion of \mntwe{} molecules on \mostwo{}.

\section*{Acknowledgments}

This work was supported by the US Department of Energy (DOE), Oﬃce of Basic Energy Sciences (BES), under contract no.{} DE-SC0022089. Computations were performed using resources of the University of Florida Research Computing as well as the National Energy Research Scientific Computing Center (NERSC), a U.S. Department of Energy Office of Science User Facility located at Lawrence Berkeley National Laboratory, operated under Contract no. DE-AC02-05CH11231.

\bibliographystyle{apsrev4-2}

%

\end{document}